\documentclass[10pt,aps,prb,twocolumn,superscriptaddress,floatfix,showpacs,longbibliography]{revtex4-1}
\usepackage[utf8]{inputenc}
\usepackage{graphicx}
\usepackage{tabularx}
\usepackage[usenames,dvipsnames,table]{xcolor}
\usepackage[version=3]{mhchem}
\usepackage{braket}
\usepackage{upgreek}
\usepackage{hyperref}
\usepackage{amsmath, amssymb}
\usepackage[normalem]{ulem}
\usepackage{soul}

\definecolor{mark}{rgb}{0.85, 0.9, 1}

\hypersetup{hidelinks}
\sethlcolor{mark}

\renewcommand{\vec}[1]{\mathbf{#1}}
\renewcommand{\phi}{\varphi}
\renewcommand{\theta}{\vartheta}

\begin{document}

\title{Exploring entanglement in finite-size quantum systems with degenerate ground state}
\author{V.S. Okatev}
\affiliation{Theoretical Physics and Applied Mathematics Department, Ural Federal University, Ekaterinburg 620002, Russia}

\author{Oleg M. Sotnikov}
\affiliation{Theoretical Physics and Applied Mathematics Department, Ural Federal University, Ekaterinburg 620002, Russia}
\affiliation{Russian Quantum Center, Skolkovo, Moscow 121205, Russia}

\author{V.V. Mazurenko}
\affiliation{Theoretical Physics and Applied Mathematics Department, Ural Federal University, Ekaterinburg 620002, Russia}
\affiliation{Russian Quantum Center, Skolkovo, Moscow 121205, Russia}

\begin{abstract}
We develop an approach for characterizing non-local quantum correlations in spin systems with exactly or nearly degenerate ground states. Starting with linearly independent degenerate eigenfunctions calculated with exact diagonalization we generate a finite set of their random linear combinations with Haar measure, which guarantees that these combinations are uniformly distributed in the space spanned by the initial eigenstates. Estimating the von Neumann entropy of the random wave functions helps to reveal previously unknown features of the quantum correlations in the phases with degeneracy of the ground state. For instance, spin spiral phase of the quantum magnet with Dzyaloshinskii-Moriya interaction is characterized by the enhancement of the entanglement entropy, which can be qualitatively explained by the changes in behaviour of two- and three-spin correlation functions. To establish the connection between our theoretical findings and real experiments we elaborate on the problem of estimating observables on the basis of the single-shot measurements of numerous degenerate eigenstates. By the example of solving a simple Ising model it is shown that digital quantum simulations performed with quantum computers can provide accurate description of the entanglement of the degenerate systems even in the presence of noise. We also discuss the ground state properties of degenerate fermionic and bosonic models, which can be useful for theoretical analysis of the experiments with ultracold atoms.
\end{abstract}

\maketitle

\section{Introduction}

In quantum mechanics, if two or more different eigenfunctions of a Hamiltonian operator that describes a quantum system correspond to the same eigenenergy, they are degenerate and their arbitrary linear combination is likewise an eigenstate of the system. Degenerate quantum systems represent a special interest in condensed matter physics, theory of magnetism and quantum computing. For instance, the notable quantum Ising model characterized by two-fold degenerate ground state is widely used for benchmarking quantum algorithms \cite{Ising_dissipation, Heyl, Carleo} and platforms \cite{Rydberg1, Rydberg2} as well as for demonstrating quantum priority \cite{IBM127}. Another important example is the Kitaev's toric code, degenerate quantum states employed in error correction algorithms \cite{Toric_code}. From the side of fundamental research, the key role of degeneracy has been revealed when explaining the formation of a classical long-range order in a quantum system in the thermodynamic limit\cite{Anderson_towers, Lauchli_towers, Quantum_Darwinism}.  

\begin{figure}[t]
  \centering
  \includegraphics[width=1\linewidth]{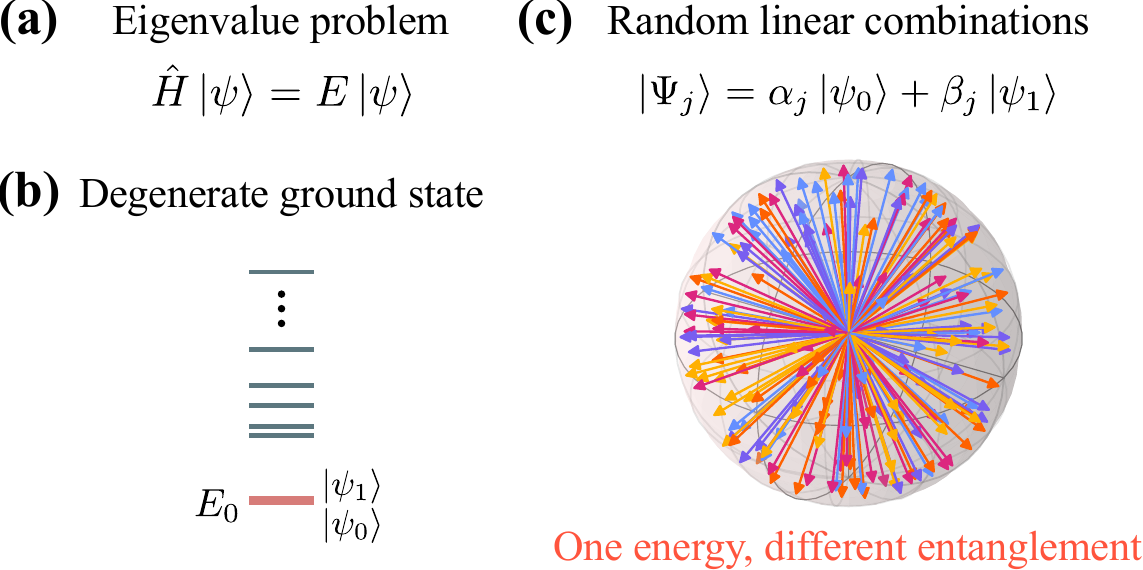}
  \caption{The idea of our study. (a) We consider the eigenvalue problem for a given Hamiltonian with degenerate ground state that can be solved by means of exact diagonalization. (b) Schematic representation of an eigenspectrum of a system with two-fold degenerate ground state. (c) Random linear combinations of the degenerate eigenfunctions correspond to the same energy, but they can be characterized by different amount of the entanglement. Exploring quantum correlations of such a random eigenstates ensemble represents the main interest for us in this work.}
  \label{intro}
\end{figure}

Exploring the properties of the quantum systems with degenerate ground state is a hard numerical problem and is mainly performed with the family of the exact diagonalization methods having fundamental limitations on the size of the simulated systems \cite{Lauchli1, Lauchli2}. Furthermore, some of the exact diagonalization approaches cannot resolve degeneracy correctly \cite{Fehske, Manmana}. Employing neural quantum state approaches allows to work around the size limitation of the exact diagonalization and find the ground state energies of larger quantum systems. However, the accurate estimation of an observable in the case of a degenerate system generally represents a vast problem for neural-based approaches\cite{Joshi}. Linear combinations of the degenerate eigenstates obtained with exact diagonalization correspond to the same energy (Fig.\ref{intro}), but they can be of different complexity with respect to the amount of entanglement \cite{Quantum_Darwinism}, phase structure \cite{Bagrov1, Bagrov2} and other quantum state properties \cite{Sotnikov}. This means that the theoretical description of a quantum system with degeneracy of the ground state is associated with the characterization of an ensemble of degenerate eigenfunctions that can be generated on the basis of the exact diagonalization solutions (Fig.\ref{intro}). 

In this sense, interesting results concerning exactly solvable quantum spin models with degeneracy were obtained in the previous works \cite{Popkov1, Popkov2}. By using the degenerate eigenstates of the one-dimensional ferromagnetic Heisenberg model the authors of Ref.\onlinecite{Popkov1} constructed density matrices of pure and mixed types and define the scaling of the entanglement entropy on the subsystem size. A distinct approach for describing entropy of the ferromagnetic Heisenberg model with highly-degenerate ground state was developed in Refs.\onlinecite{Doyon1, Doyon2, Doyon3}, where it was shown that any eigenfunction from degenerate manifold of the $n$-qubit system can be represented as a linear combination of the $n+1$ trivial zero-entropy permutation symmetric states. On this basis the authors explored entanglement behaviour of the ferromagnetic Heisenberg model in the thermodynamics limit. To our knowledge, the entanglement properties of degenerate quantum systems beyond above-mentioned cases of the integrable models has been practically unexplored. 

From the experimental side, each preparation of the ground state of a degenerate quantum system with adiabatic \cite{Rydberg1} or dissipative \cite{Ising_dissipation} protocols can result in a distinct wave function from degenerate eigensubspace. This should be taken into account when estimating different correlation functions on the basis of the projective measurements, which, in general, requires multiple copies of the same wave function. 

In this paper we address the problem of characterizing degenerate ground states by the examples of the one-dimensional quantum Ising model and a two-dimensional spin system with Heisenberg and Dzyaloshinskii-Moriya interactions. Both models are characterized by the same mechanism of changing the degeneracy of the ground state that is related to external magnetic field. First, to explore the ground state properties of these systems we generate finite sets of wave functions that belong to the degenerate eigensubspace and calculate the distributions of their von Neumann entropies. Remarkably, in the case of the model with non-collinear magnetic order, a previously unknown enhancement of the entanglement entropy of the degenerate states ensemble within the spin-spiral phase has been revealed. We show that such an enhancement correlates with change in behaviour of the spin-spin correlation functions at the same magnetic field values. To establish a connection with previous works \cite{Popkov1, Popkov2,Doyon1, Doyon2, Doyon3} we analyze the case of the highly-degenerate ground states of the one-dimensional ferromagnetic Heisenberg model. Likewise, we consider the problem of reconstructing the simplest local spin correlation function by using the results of the projective measurements in the extreme case when before each measurement the system in question is prepared in a unrepeatable state from the degenerate manifold, which is relevant to the conditions of the real experiments.    

\section{Two-fold degenerate ground state of Ising model}
We start our investigation with analyzing the paradigmatic  model in condensed matter physics that is transverse-field Ising Hamiltonian. Such a model for which the exact solution\cite{Pfeuty} is known since 1970 is still actively used for certifying various quantum algorithms \cite{Carleo, Heyl, Sotnikov, Fernandez, Ising_QAOA}, testing utility of quantum computation with different platforms \cite{Rydberg1, Rydberg2, IBM127}, exploring non-equilibrium phases of matter \cite{DTC_PRX, DTC_Google, ourDTC} and solving other problems.  The Ising Hamiltonian is given by
\begin{eqnarray}
\label{Ising_model}
{\hat H}_{\rm Ising} = \sum_{ij} J_{ij} {\hat S}^z_{i} {\hat S}^z_{j} + h \sum_{i} {\hat S}^x_{i},
\end{eqnarray}
where ${\hat S}^z_{i}$ and ${\hat S}^x_{i}$ are the spin-$\frac{1}{2}$ operators, $J_{ij}$ is the ferromagnetic exchange interaction between nearest neighbours along the chain and $h$ is the $x$-oriented magnetic field. The considered system is characterized by the periodic boundary conditions. According to Ref.\onlinecite{Pfeuty} such a model features transition from ferromagnetic to paramagnetic states at the critical magnetic field value of 0.5$J$. 

\begin{figure}[b]
  \centering
  \includegraphics[width=1\linewidth]{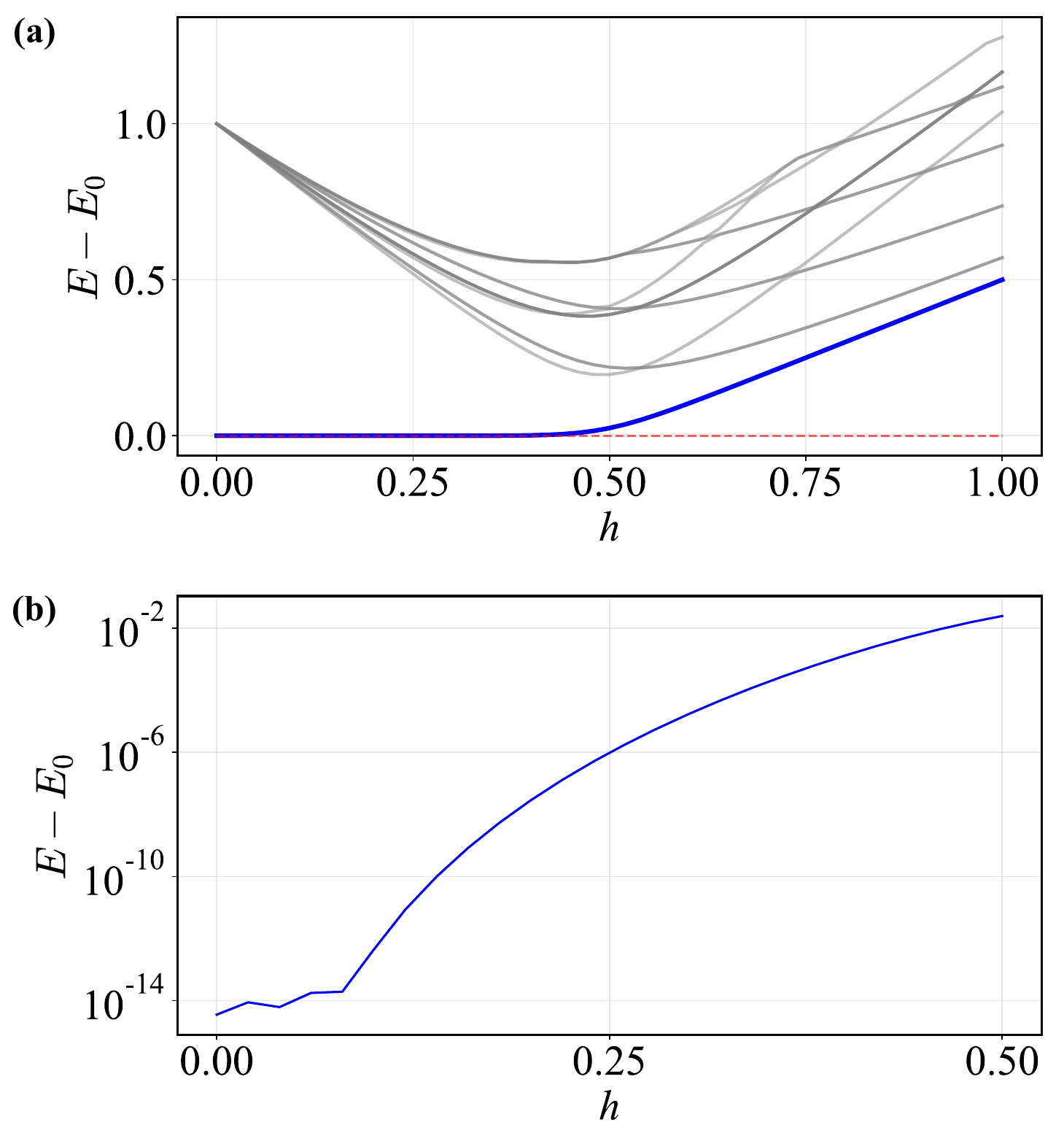}
  \caption{(a) The calculated energies of the 16 low-lying eigenstates of the 16-spin Ising model in the transverse magnetic field, Eq.\ref{Ising_model}. At each magnetic field the energies are shifted with respect to the ground state one (red dashed line). (b) Energy gap between ground and first excited states obtained for the ferromagnetic phase of the transverse-field Ising model. At $h=0.5$ the energy gap is equal to 0.0246.}
  \label{energy_Ising}
\end{figure}

Figure~\ref{energy_Ising} gives the lowest energy part of the eigenspectrum of the 16-spin transverse-field Ising model obtained from exact diagonalization \cite{Tom} with the nearest neighbour interaction $J = -1$. At the zero magnetic field the ground state of the Ising model is two-fold degenerate. Turning on the magnetic field leads to the opening of a gap between ground and first excited states, at the same time, the magnitude of the gap is of order of machine precision for weak fields as can be seen from Fig.\ref{energy_Ising}. Increasing the magnetic field gradually enlarges the energy gap and at the critical point one has $\Delta$ = 0.0246 which is an order of magnitude smaller than the splitting between first and second excited states at the same field. From the perspective of realizing the Ising model in real experiments at finite temperatures, for instance with Rydberg atoms \cite{Rydberg} and trapped ions \cite{ions},  the ground and first excited states of the Ising model within the ferromagnetic phase can be safely considered as being nearly degenerate.     

\begin{figure}[b]
  \centering
  \includegraphics[width=1\linewidth]{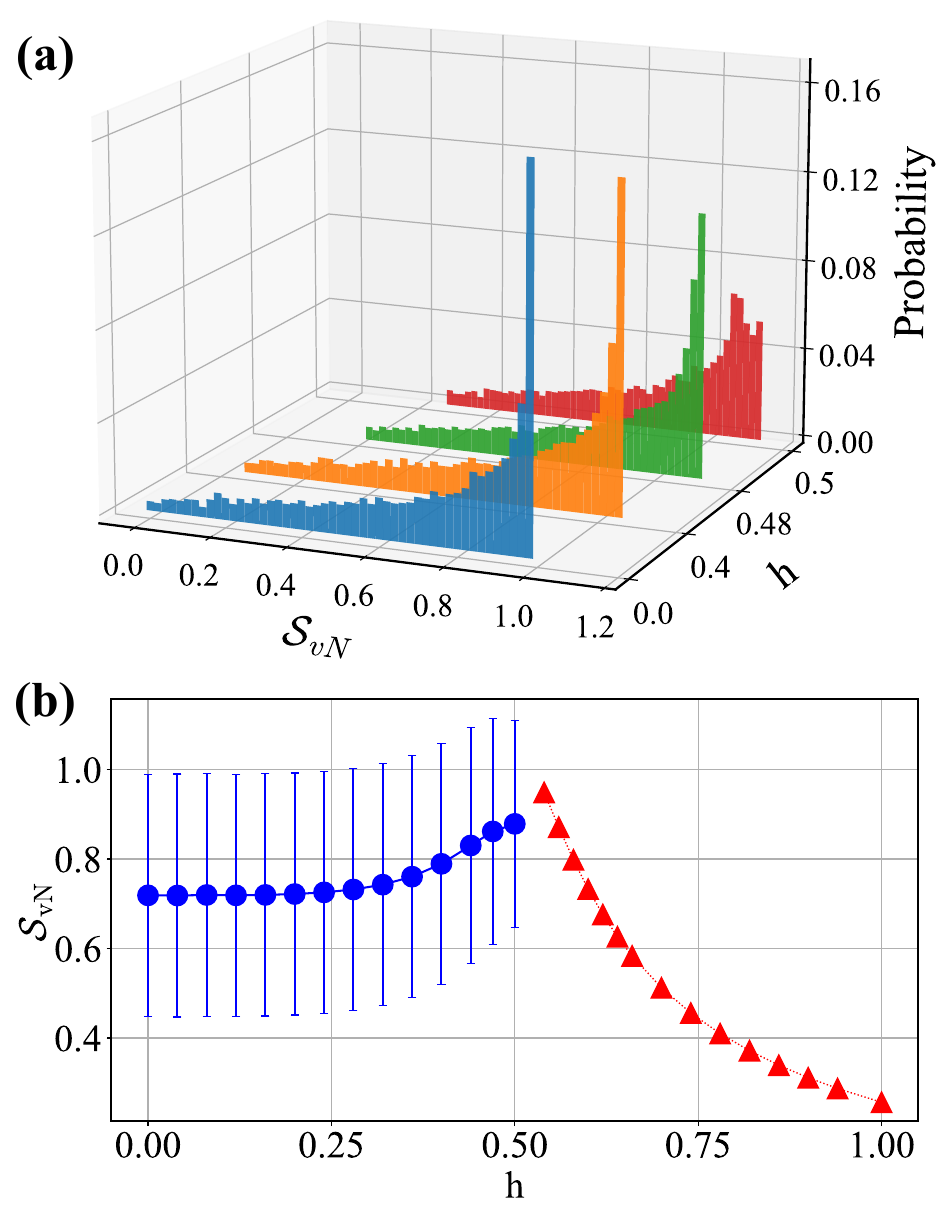}
  \caption{ (a) Distributions of the von Neumann entropy values calculated for 8192 samples at the different magnetic fields corresponding to the degenerate regime of the one-dimensional Ising model. (b) Dependence of the von Neumann entropy calculated for 16-qubit Ising model on the magnetic field. Error bar for the results obtained at $h \in [0, 0.5]$ denotes the standard deviations of the entropies of the 8192 eigenstates generated by using Haar measure for degenerate ground states as described in the text.}
  \label{svn}
\end{figure}

Now we will turn to the analysis of the entanglement entropy that is of special interest in the case of the Ising model due to the possibility of analytical treatment \cite{entropy_Ising1,entropy_Ising2}.  
The gist of the problem we are going to solve in this paper can be demonstrated by the example of the zero magnetic field Ising model for which the ground states of $n$-spin system are the trivial wave functions, $\Ket{\phi_{1}} = \Ket{\uparrow}^{\otimes n}$ and  $\Ket{\phi_{2}} = \Ket{\downarrow}^{\otimes n}$. Since each wave function is the tensor product of states of individual spins, they are non-entangled ones and the von Neumann entropy calculated for the reduced density matrix of a subsystem A, $\mathcal{S}_{\rm vN} (\rho_{\rm A}) = - {\rm Tr} \rho_{\rm A} \log_2 \rho_{\rm A}$ is zero. We would like to stress that one can choose a different set of wave functions  $\Ket{\phi_{+}} = \frac{1}{\sqrt{2}}(\Ket{\uparrow}^{\otimes n} + \Ket{\downarrow}^{\otimes n})$ and  $\Ket{\phi_{-}} = \frac{1}{\sqrt{2}}(\Ket{\uparrow}^{\otimes n} - \Ket{\downarrow}^{\otimes n})$ to define the two-fold degenerate ground state of the Ising model. In this case we deal with more complex wave functions whose entanglement entropy $\mathcal{S}_{\rm vN} (\rho_{\rm A}) = 1$ for any subsystem A. 

This simple example with different choices of the eigenstates clearly shows that entanglement properties of the degenerate ground states can be very diverse, as likewise found in the DMRG study of the Ising model \cite{DMRG}. From our point of view, a complete theoretical description of the entanglement entropy of the degenerate eigenstates should be based on consideration of their ensemble and the main question is how to generate such an ensemble. Put another way, one is to specify the way of generating the complex random coefficients, $\alpha_{0}, \alpha_{2}, ... \alpha_{\mathcal{D}-1}$ (where $\mathcal{D}$ is the degeneracy degree)  for linear combinations of the linearly independent degenerate wave functions, $\{ \Ket{\psi_{d}} \}_{d = 0.. \mathcal{D}-1}$ obtained from exact diagonalization. Assuming that the resulting eigenstates should be uniformly distributed in the state space spanned with the wave functions $\{ \Ket{\psi_{d}} \}_{d = 0.. \mathcal{D}-1}$, the real and imaginary parts of the coefficients $\{ \alpha_{d} \}_{d=0..\mathcal{D}-1}$ are independently chosen from the normal distribution with zero mean \cite{Haar_qiskit} and, then, thus obtained vector is normalized. As we will show below such a procedure provides invariance of the ensemble properties to the choice of the initial eigenvectors, $\{ \Ket{\psi_{d}} \}_{d = 0.. \mathcal{D}-1}$. In this work the degeneracy degrees are $\mathcal{D} = 2$ (Ising model) and $\mathcal{D} = 6$ (DMI model), which defines the dimensions of the corresponding eigenvectors subspaces.

Figure~\ref{svn}\,(a) gives the entanglement distribution calculated for the following linear combinations of the degenerate eigenstates of the 16-spin Ising model
\begin{eqnarray}
\label{superposition}
\Ket{\Psi} = \alpha_{0} \ket{\psi_{0}} + \alpha_1 \ket{\psi_{1}}.
\end{eqnarray}
Here $\ket{\psi_0}$ and $\ket{\psi_1}$ are the ground eigenstates taken from the exact diagonalization, and the random complex coefficients $\alpha_0$  and $\alpha_1$ are generated by using the Gaussian distribution as described above. Having generated 8192 different eigenstates $\ket{\Psi}$ we have found that the probability of the samples with zero or small entanglement is close to zero. At the same time, ${\rm Pr}(S_{\rm vN})$ grows as the entanglement increases and achieves the maximal value at $S_{\rm vN} (\rho_{A}) = 1$.
 As the result, the mean of the entanglement distribution is located at 0.75 and the corresponding standard deviation mainly covers the area of high entanglement (Fig.~\ref{svn}\,(b)).

Numerical simulations of the ensembles each of which consists of 8192 eigenfunctions generated at finite magnetic fields with Eq.\ref{superposition} reveal the entanglement distributions similar to that for the zero-field case. For $h < 0.25$ we observe a constant behaviour of the mean value of the calculated von Neumann entropy. Importantly, in the vicinity of the critical point $h = 0.5$ there is an enhancement of the bi-partite entanglement mean value. This agrees well with the results of the previous analytical considerations of the Ising model \cite{entropy_Ising1,entropy_Ising2} and can be explained by analyzing nonlocal spin-spin correlations, observable quantities which detect the presence of entanglement in a system \cite{Entanglement_witnesses1, Entanglement_witnesses2}. According to the results of the seminal work by Pfeuty (Table I, Figs. 4 and 5 in Ref.\onlinecite{Pfeuty}) the spin-spin correlation functions diverge at the critical field and are characterized by a distinct decay with increasing the distance between spins.  As we will show below other quantum systems with degenerate eigenstates can likewise feature enhancement of the entanglement at varying external parameters.
At $h > 0.5$ the entanglement entropy has been calculated for non-degenerate ground states and demonstrates gradual decrease as the magnetic field increases. It approaches to zero in the limit of $h \rightarrow \infty$.

To demonstrate the importance of using the Gaussian distribution for sampling random coefficients in Eq.\ref{superposition} which corresponds to the Haar measure and provides the invariance of the ensemble properties with respect to the choice of the initial eigenstates, we consider two types of ground states superpositions:
\begin{equation}
  \label{eq:ising_random}
  \begin{split}
    \ket{\Psi}_d &= \alpha_0 \Ket{\uparrow}^{\otimes n} + \alpha_1 \Ket{\downarrow}^{\otimes n}, \\
    \ket{\Psi}_e &= \frac{\alpha_0}{\sqrt{2}}(\Ket{\uparrow}^{\otimes n} + \Ket{\downarrow}^{\otimes n}) + \frac{\alpha_1}{\sqrt{2}}(\Ket{\uparrow}^{\otimes n} - \Ket{\downarrow}^{\otimes n}),
  \end{split}
\end{equation}
where $\alpha_0$ and $\alpha_1$ are random coefficients that can be generated with some distributions. The second equation can be rewritten as 
\begin{equation}
  \ket{\Psi}_e = \frac{\alpha_0+\alpha_1}{\sqrt{2}}\Ket{\uparrow}^{\otimes n} + \frac{\alpha_0-\alpha_1}{\sqrt{2}}\Ket{\downarrow}^{\otimes n}.
\end{equation}
Then, one can express the von Neumann entropy for these superpositions as
\begin{equation}
  \begin{split}
\label{entropy_Ising}
    \mathcal{S}_\text{vN}^d (\rho_{A}) &= -|\alpha_0|^2\log_2|\alpha_0|^2 -|\alpha_1|^2\log_2|\alpha_1|^2, \\
    \mathcal{S}_\text{vN}^e (\rho_A) &= -\frac{|\alpha_0+\alpha_1|^2}{2}\log_2\frac{|\alpha_0+\alpha_1|^2}{2} \\
    &\quad -\frac{|\alpha_0-\alpha_1|^2}{2}\log_2\frac{|\alpha_0-\alpha_1|^2}{2},
  \end{split}
\end{equation}
where $\rho_A$ is the reduced density matrix describing the subsystem A. With the expressions, Eq.\ref{entropy_Ising} it is possible to sample entropy of arbitrary size system for different distribution laws of $\alpha_0$ and $\alpha_1$ random variables. 

\begin{figure}
  \centering
  \includegraphics[width=1\linewidth]{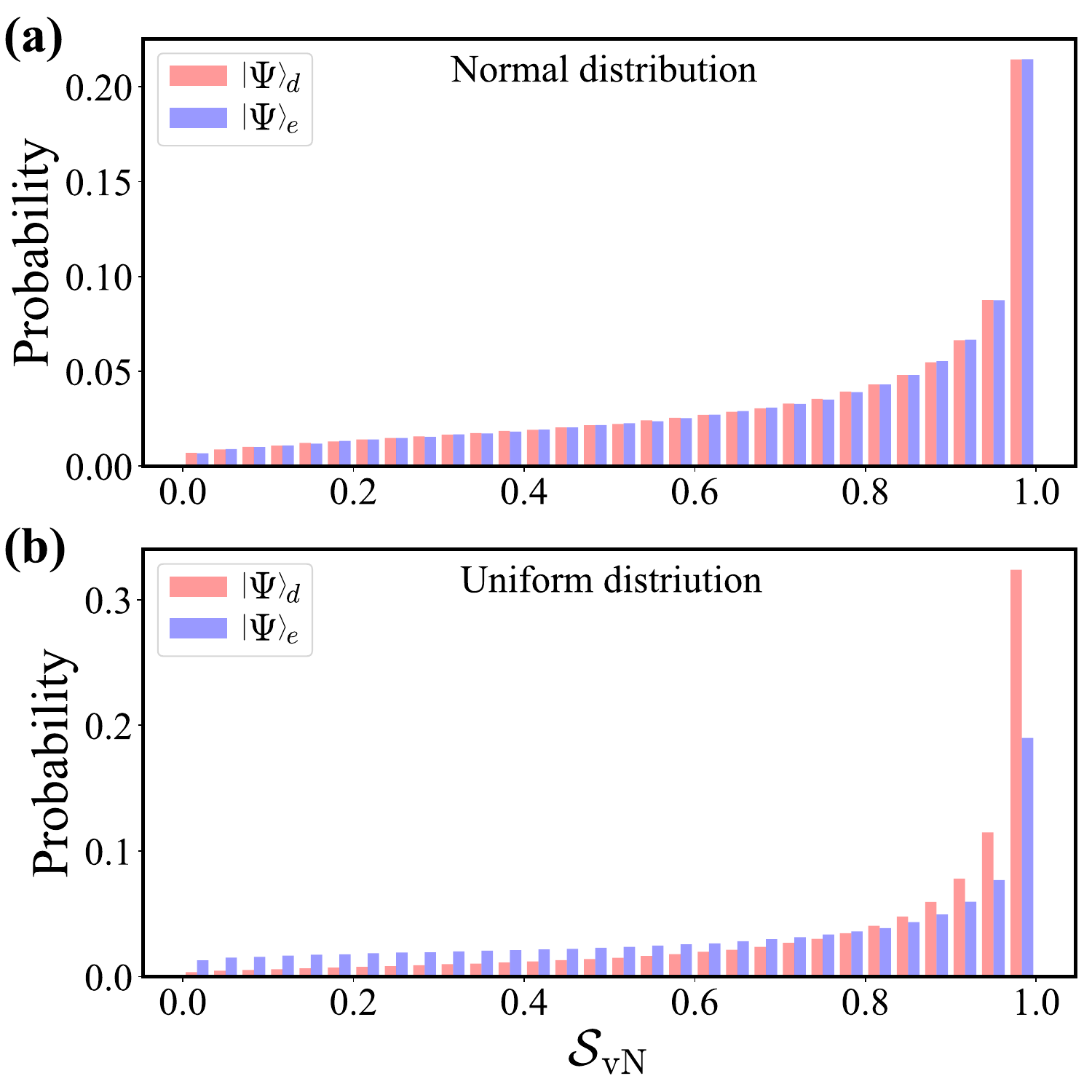}
  \caption{Entropy distributions calculated for the superpositions of the entangled $\Ket{\Psi_e}$ and non-entangled $\Ket{\Psi_d}$ ground states of the zero-field Ising model. Data were obtained with (a) normal (Haar) and (b) uniform distributions for the linear combination coefficients.}
  \label{fig:ising_entanglement}
\end{figure}

Figure~\ref{fig:ising_entanglement} demonstrates the examples of the entropy distributions for $\ket{\Psi}_d$ and $\ket{\Psi}_e$ states when the real and imaginary parts of the coefficients are generated by using either uniform distribution in range (-1,1) or Gaussian distribution with zero mean and unit variance. While for the Gaussian case we observe invariance of the resulting entanglement distribution with respect to the initial states used for ensemble creation (Fig.~\ref{fig:ising_entanglement}\,(a)), it is not the case for the results obtained with the uniform distribution (Fig.~\ref{fig:ising_entanglement}\,(b)). 

\subsection{Impact of decoherence}
Since in real experiments the quantum systems of interest can never be isolated from their environment and experimental measurements are imperfect themselves, thus, it is important to elaborate on the impact of decoherence on entanglement of degenerate ground states we study. In experiments aimed at imitating a spin model such as Ising model \cite{Ising_dissipation, Rydberg1, Rydberg2} the ground wave functions are not given by default, they should be first prepared and then measured. Obviously, decoherence should be taken into account at all the stages of the preparation procedure and measurement as well. Here one should distinguish analog and digital quantum simulations, since they assume different amount of information about degenerate system in question that can be extracted from the measurements. When for imitating a spin model with degeneracy of the ground state one uses a physical system with tuned interactions (analog simulations), for instance assembly of Rydberg atoms, this is the case of maximal uncertainty and information deficit, since a distinct wave function from degenerate eigensubspace is prepared and measured each time. Thus, having performed a number of such experiments one actually has information about different wave functions. We will consider such a scenario in the section V.

\begin{figure}[!b]
  \centering
  \includegraphics[width=1\linewidth]{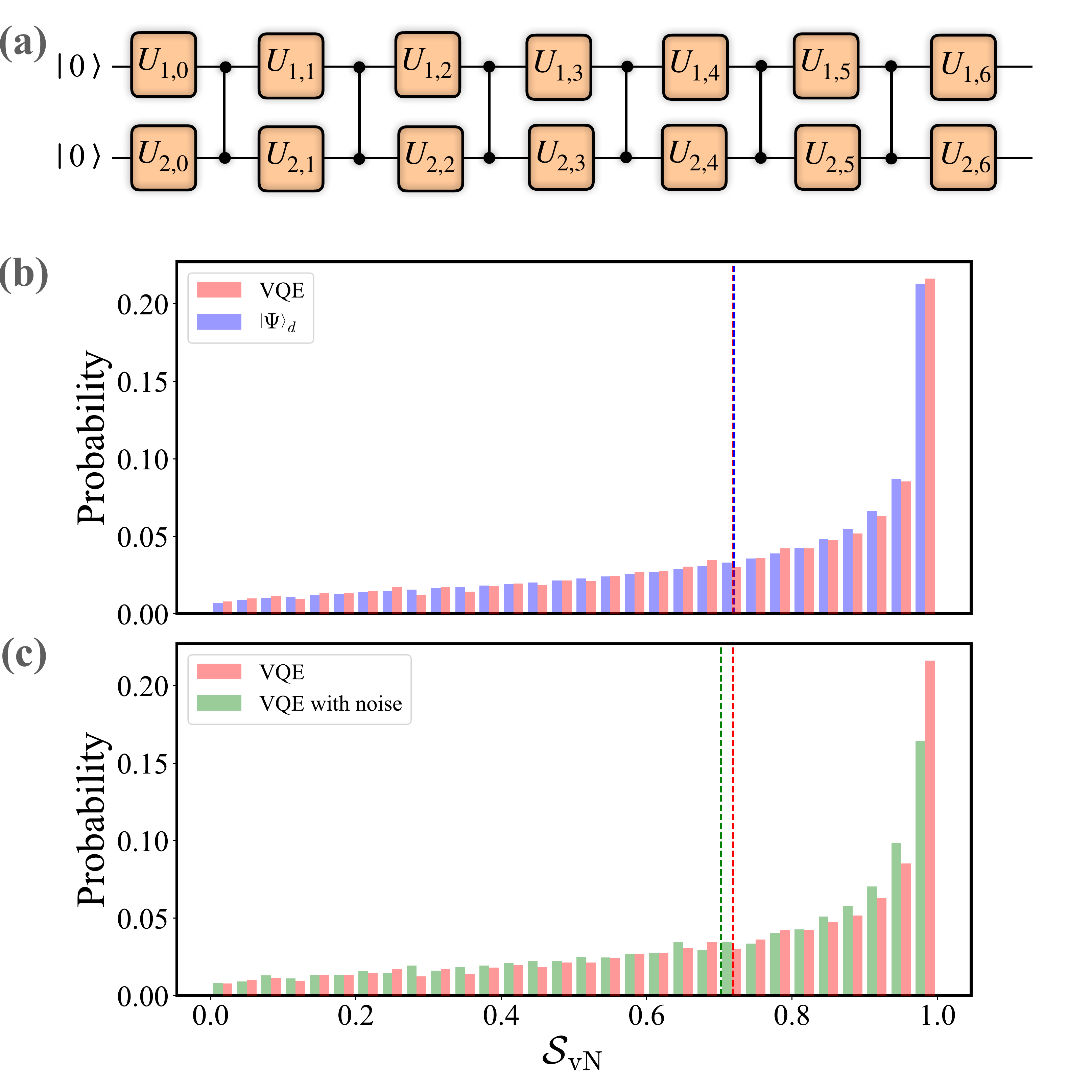}
  \caption{(a) Schematic representation of the quantum circuit used in VQE simulations of two-spin Ising model with degenerate ground state. Each single-qubit unitary operation, $U_{i,d}$ is defined with three independent angles.  (b) Comparison of the entanglement probability distributions obtained with 8192 VQE noise-free solutions (red) and calculated for 8192 linear combinations $\ket{\Psi}_d$ sampled with Haar measure, Eq.\ref{eq:ising_random} (blue). (c) Comparison of the entanglement distributions of the noisy (green) and noise-free (red) VQE solutions. The dashed lines denote the mean values of the presented entanglement distributions.}
  \label{VQE}
\end{figure}

Contrary to the analog quantum simulations described above, the digital ones allow more accurate description of the degenerate eigenstates. This is due to the fact that, having found an eigenstate from degenerate manifold one can reproduce it as many times as necessary and thus perform numerous measurements, which paves the way for complete characterization including quantum state tomography. To demonstrate this we solve the simplest zero-field two-site ferromagnetic Ising model, Eq.\ref{Ising_model} on a quantum simulator with the variational quantum eigensolver \cite{VQE} (VQE) by using the Qiskit package \cite{qiskit}. Within this approach the wave function of the target quantum system is represented as
\begin{eqnarray}
\label{ansatz}
\Ket{\psi(\boldsymbol \Theta)} = \mathcal{U}(\boldsymbol \Theta) \ket{00},
\end{eqnarray}
where the parametrized unitary transformation has the following form
\begin{eqnarray}
\label{circuit}
\mathcal{U} (\boldsymbol \Theta) = \bigg[ \prod_{d=1}^{D} \big[ \prod _{i=1}^2 U(\boldsymbol \theta_{i,d}) \big] \times U_{\rm CZ} \bigg]  \times \big[ \prod _{i=1}^2 U(\boldsymbol \theta_{i,0}) \big].
\end{eqnarray}
Here $D$ denotes the depth of the quantum circuit, $U_{\rm CZ}$ stands for two-qubit controlled-Z gate, $U(\boldsymbol \theta_{i,d})$ is the universal rotational gate acting on $i$th qubit in the $d$th layer of the quantum circuit. 
Fig.\ref{VQE} a gives the schematic visualization of the quantum circuit, Eq.\ref{circuit} used for approximating the ground state of the ferromagnetic Ising dimer model. There are 6 layers of one-qubit rotational gates and two-qubit CZ gates. Taking into account the initial rotational operators $U_{1,0}$ and $U_{2,0}$, the total number of the optimized parameters (angles) within the VQE procedure is 42.

We have performed 8192 simulations of the two-site Ising model by using the VQE approach in the decoherence-free regime. The ground state energy averaged over these experiments is about -0.2498 and the corresponding standard deviation equals to  0.0002, which means that VQE provides excellent agreement with exact result. For each thus found wave function we have calculated the von Neumann entropy. In Fig. \ref{VQE} b we show that the VQE solutions are characterized by the same distribution of the entanglement entropy as we found by sampling wave functions from degenerate manifold (the previous section ''Two-fold degenerate ground state of Ising model''). Thus we conclude that the chosen ansatz, Eq.\ref{ansatz} for approximating ground state of the Ising model allows one accurately explore the properties of the degenerate eigensubspace and produces the solution statistics according to Haar measure.

To simulate the impact of decoherence on the properties of the Ising dimer we have used a noise model that was generated to imitate the functioning of a real IBM Yorktown quantum processor. The considered noise model is characterized by one- and two-qubit errors which include depolarizing and thermal relaxation errors. The measurements are also subject to readout errors. For selected device readout error probability varies from 0.03 to 0.3 depending on specific qubit. Having averaged the results of 8192 VQE experiments taking into account the noise we have found the ground state energy of about -0.16 that is considerably higher than the exact value of -0.25. Such a discrepancy can be explained by the length of the quantum circuit and the number of the two-qubit operations. The decoherence mainly affects the high-entropy part of the entanglement distribution, which results in a small decrease of the mean value of $S_{\rm vN}$. Another distribution feature that is sensitive to decoherence is skewness which decreases from 3.73 for ideal VQE to 2.78 for noisy VQE. 

The environment-induced suppression of the probability to find a highly-entangled state from degenerate manifold is generally in line with the main idea of the decoherence theory \cite{Joos1, Zurek0,Zurek1, Zurek3, Zurek2, Quantum_Darwinism}, which assumes a selection and promotion of most trivial states by the environment so that different independent observers find them in experiments. However, it seems natural that this theory should be adopted when used in a new context like digital quantum simulations considered in this work. We leave such an integration of the approaches for further study.

\begin{figure*}
  \centering
  \includegraphics[width=1\linewidth]{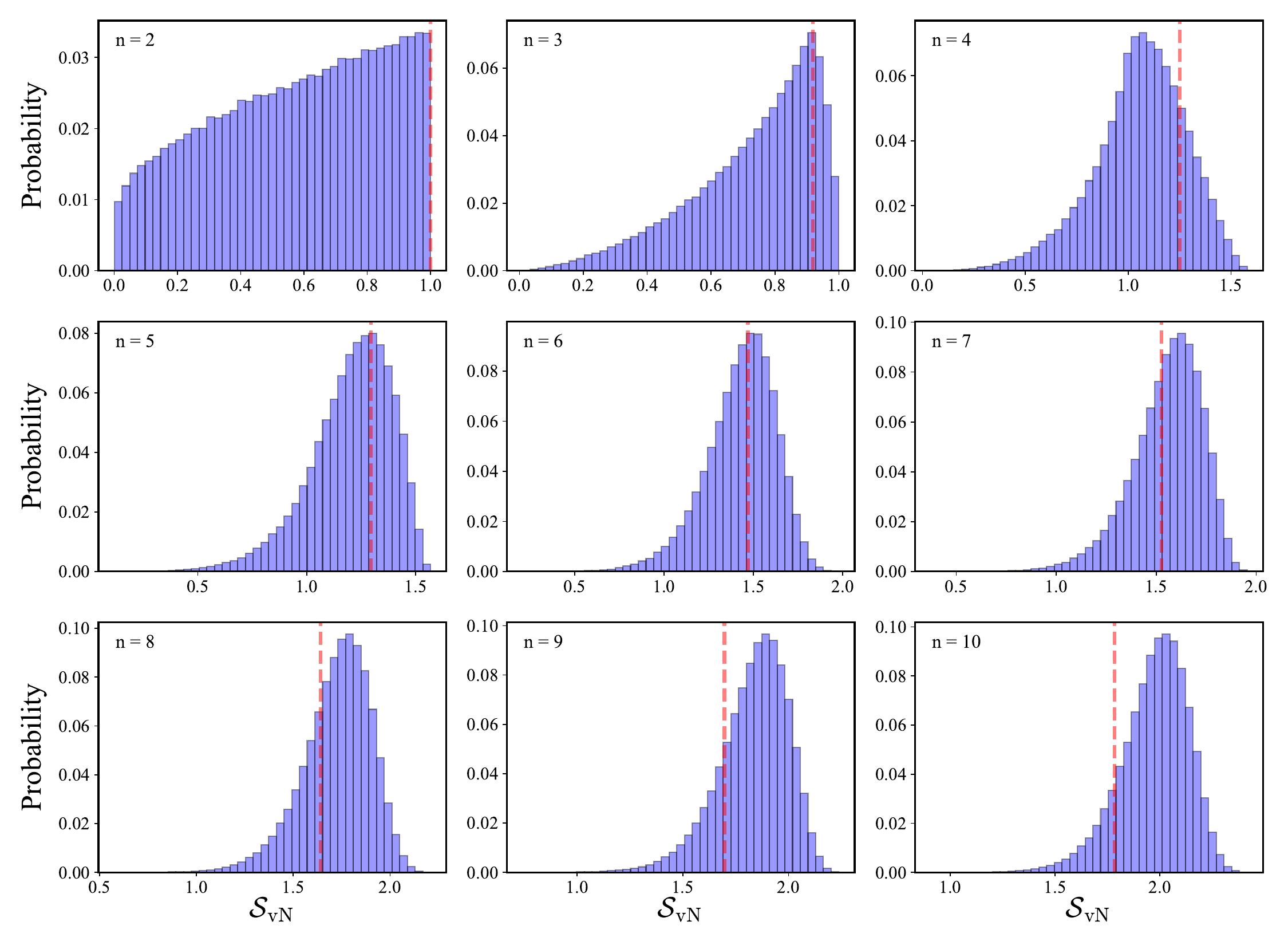}
  \caption{Entropy distributions calculated for the one-dimensional Heisenberg models of $n$ = 2,3,...10 spins of $\frac{1}{2}$. In each case $2^{18}$ random linear combinations of the ground eigenstates (Dicke wave functions) were generated. The red lines indicate the entanglement entropy of the most complex Dicke states for a given number of spins.}
  \label{fig:HeisFM}
\end{figure*}

\section{Highly-degenerate ground state of one-dimensional ferromagnetic Heisenberg model}
The one-dimensional ferromagnetic Ising model explored above represents an example of the quantum system with constant degeneracy of the ground state that is two-fold and independent on the system size. The transition from Ising to Heisenberg model, $\hat{H}_{\rm Heis} = \sum_{i>j}J_{ij}\hat{\vec{S}}_i\cdot{}\hat{\vec{S}}_j$ suggests  completely different picture of the ground state degeneracy whose degree is controlled by the size of the system. For the one-dimensional ferromagnetic spin-$\frac{1}{2}$ Heisenberg model with periodic boundary conditions, the degeneracy of the ground state scales linearly to the system size and the corresponding ground eigenfunctions are nothing but the Dicke states \cite{Popkov1, Doyon2} that are given by \cite{Dicke}
\begin{eqnarray}\label{Dicke_wf}
	\Ket{D^{d}_{n}} = \frac{1}{\sqrt{C^d_n}} \sum_{j} P_{j}(\Ket{\uparrow}^{\otimes n-d} \otimes \Ket{\downarrow}^{\otimes d}).
\end{eqnarray}
Here the sum goes over all possible permutations $P_j$ of spins, and $C_n^d$ is the number of $d$-combinations from a set of $n$ elements. For the Heisenberg system of $n$ interacting spins, the degenerate ground eigensubspace is spanned by $n+1$ Dicke  wave functions, $\{ \Ket{D^{0}_{n}}$, $\Ket{D^{1}_{n}}$, $\Ket{D^{2}_{n}}$,..$\Ket{D^{n}_{n}} \}$. Importantly, these Dicke states are of different complexity. For instance, $\Ket{D^{0}_{n}}$ and $\Ket{D^{n}_{n}}$ are trivial zero-entropy wave functions and $\Ket{D^{[\frac{n}{2}]}_{n}}$ is characterized by quantum correlations at any order of a coarse grained partition \cite{Dicke_correlations}. Based on this, one can expect to observe interesting entanglement properties when generating wave functions from eigensubspace with random linear combinations $\ket{\Psi} = \sum_{d=0}^{n} \alpha_{d} \Ket{D^{d}_{n}}$ for different sizes of the system in question. Indeed, the results presented in Fig.\ref{fig:HeisFM} reveal completely different patterns of the von Neumann entropy entanglement for the Heisenberg models with $n$ = 2, 3, 4 and 5. For larger systems the entropy distribution is characterized by a Gaussian-like profile whose mean gradually shifts to higher  values as increasing the number of sites.

Another interesting conclusion follows from comparison (Fig.\ref{fig:HeisFM}) of the mean entropy value averaged over ensemble of the degenerate states for the $n$-spin system and those obtained for the most complex Dicke states for the given system size. Previous studies have revealed a logarithmic scaling of the von Neumann entropy on the system size. Fig.\ref{fig:HeisFM} clearly shows that the mean entropies of the linear combinations grow faster than $\mathcal{S}_{\rm vN}$ calculated for the most complex Dicke states. In addition, the dispersion of the ensemble entropies decreases with increasing $n$. 

Since the degeneracy of the ground state of the considered $n$-spin Heisenberg model is equal to $n+1$, then the degenerate eigensubspace consisting of the Dicke wave functions becomes more and more complex as the number of spins increases. Potentially, the consideration could be simplified by using the approach developed in the works \cite{Doyon1,Doyon2, Doyon3}. Namely, any Dicke state can be expressed as a linear superposition of the trivial wave functions where each function features all spins point in the same direction $\bf v$ on the Bloch sphere, $\ket{\psi_{\bf v}} = \ket{\bf v}^{\otimes n}$. For example, in the two-spin case one can define $\Ket{D^{0}_{2}} = \ket{\uparrow \uparrow}$, $\Ket{D^{1}_{2}} = \sqrt{2} \ket{\rightarrow \rightarrow} - \frac{1}{\sqrt{2}} \ket{\uparrow \uparrow} -\frac{1}{\sqrt{2}} \ket{\downarrow \downarrow}$ and $\Ket{D^{2}_{2}} = \ket{\downarrow \downarrow}$, where $\ket{\rightarrow} = \frac{1}{\sqrt{2}} (\ket{\uparrow}+\ket{\downarrow})$. Thus, instead of constructing the degenerate eigensubspace of the Heisenberg model by using the three Dicke states of different complexity, one can perform the same task with zero-entropy permutationally invariant wave functions, $\ket{\uparrow \uparrow}$, $\ket{\rightarrow \rightarrow}$ and $\ket{\downarrow \downarrow}$. Put another way, any ground state from degenerate manifold of the one-dimensional ferromagnetic Heisenberg model has support on the Bloch sphere.

Simplification of the degenerate basis support described above complicates the procedure of generating finite ensemble of the wave functions that should be uniformly distributed over degenerate eigensubspace. On the level of the the Dicke states which are orthogonal to each other, one can generate each coefficient in a linear superposition independently on others by the normal distribution as we discuss in the section II. At the same time, for the quantum Heisenberg model of a finite size the trivial functions, $\ket{\psi_{\bf v}}$ supported on the Bloch sphere are not mutually orthogonal. For the two-site example, the corresponding overlaps are given by $\braket{\uparrow \uparrow |\rightarrow \rightarrow} = \braket{\downarrow \downarrow |\rightarrow \rightarrow } = \frac{1}{2}$. Thus, the particular structure of the overlaps depending on the choice of the $\ket{\psi_{\bf v}}$ vectors should be taken into account when generating coefficients of the linear superpositions. It is worth noting that the overlap between two arbitrary $\ket{\psi_{\bf v}}$ and $\ket{\psi_{\bf u}}$ vectors decreases as the number of spins increases and becomes zero in the thermodynamics limit. The entanglement behaviour of the ferromagnetic Heisenberg model in the $n \rightarrow \infty$ limit  has been studied in detail in Refs.\cite{Doyon1,Doyon2, Doyon3}.

\section{Quantum model with non-collinear magnetic ordering}
As an example of the non-integrable quantum system with $\mathcal{D} > 2$ we consider the Heisenberg Hamiltonian with Dzyaloshinskii-Moriya interaction defined on a 19-site triangular supercell with periodic boundary conditions:
\begin{equation}
\label{DMI_Ham}
  \hat{H}_{\rm DMI} = \sum_{i>j} J_{ij}\hat{\vec{S}}_i\cdot{}\hat{\vec{S}}_j + \sum_{i>j}\vec{D}_{ij}\cdot[\hat{\vec{S}}_i\times\hat{\vec{S}}_j] + h\sum_i\hat{S}^z_i
\end{equation}
where $J_{ij}$ and  $\vec{D}_{ij}$ are isotropic and  Dzyaloshinskii-Moriya interactions. In the previous works \cite{qskyrmion, Quantum_Darwinism} it has been shown that depending on the value of the magnetic field such a model taken with the ferromagnetic exchange $J = -0.5$ and in-plane Dzyaloshinskii-Moriya interactions $|\vec{D}_{ij}| = 1$ can be in either collinear ferromagnetic or non-collinear, quantum spin spiral or quantum skyrmion states. The origin of these quantum phases has been established by analyzing the calculated magnetization, spin structural factors, scalar chirality as well as by comparing quantum and classical solutions. Here we primarily focus on a still poorly understood issue, which concerns the entanglement properties of the spin spiral phase characterized by degeneracy of ground eigenstates. 

\begin{figure}
  \centering
  \includegraphics[width=1\linewidth]{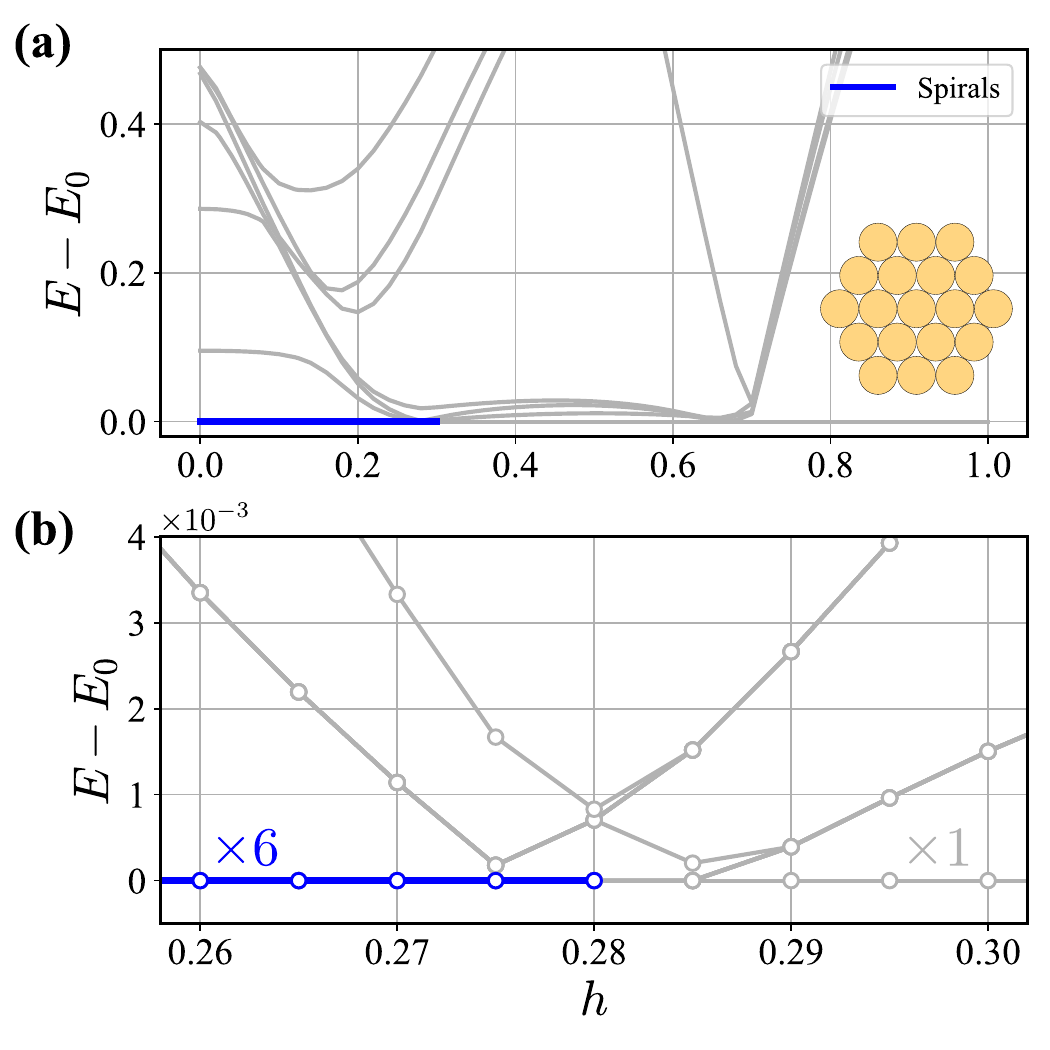}
  \caption{(a) Magnetic-field dependence of the low-energy part of the eigenspectrum calculated for the Heisenberg model with Dzyaloshinskii-Moriya interaction, Eq.\ref{DMI_Ham}. The model is defined on the 19-spin plaquette (inset). (b) Zoom view of the transitional area between the spin-spiral and skyrmion phases. Blue line denotes 6-fold degenerate spin-spiral ground state.}
  \label{fig:flake_spectra}
\end{figure}%

According to the eigenspectra obtained from exact diagonalization and presented in Fig.~\ref{fig:flake_spectra}  there is the sixfold degenerate ground state ($0 \leq h \leq 0.28$) that is denoted with blue, associated to spin spiral state and represents the main interest for us in this work.  In turn, the degeneracy is lifted for $h \in (0.28, 0.66)$.  Then, there is a very narrow range $[0.66, 0.68]$ that is characterized by the six-fold degenerate ground state. For $h > 0.68$ the ground state is non-degenerate. Taking into account the behaviour of the excited states one can distinguish two different regimes for $0.28< h< 0.68$ and $h> 0.68$. For the former 18 excited states having the energies that are close to the ground state are well separated from the rest and, thus they can be considered to be a part of nearly 19-fold degenerate ground state. This point was discussed in details in Ref.~\cite{Quantum_Darwinism}. In turn, the high field region is described by the pure ferromagnetic ground state that is well-separated from excited levels. 

\begin{figure}[!t]
  \centering
  \includegraphics[width=\linewidth]{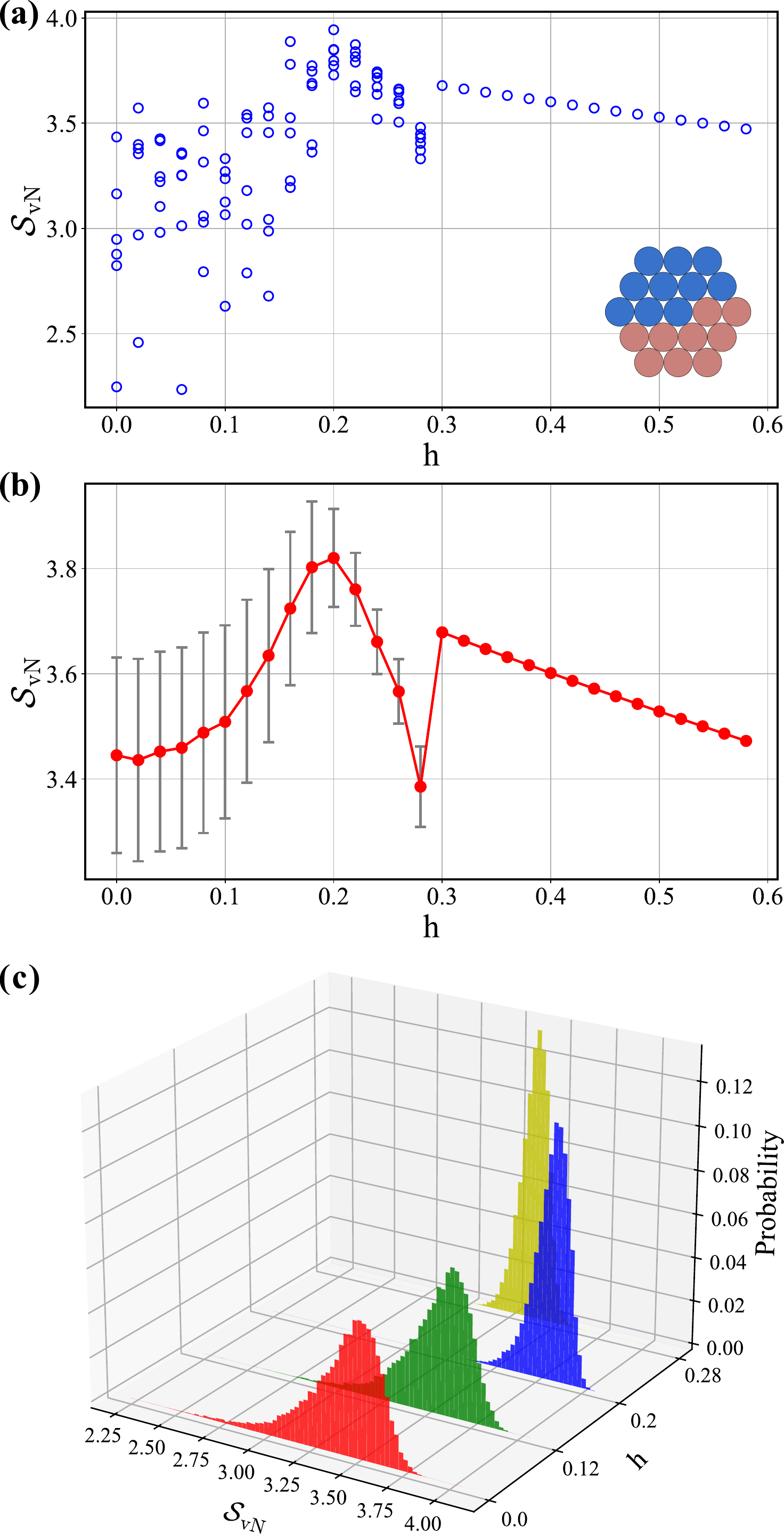}
  \caption{(a) Entanglement entropy of the eigenfunctions obtained by using the exact diagonalization of the DMI-model, Eq.\ref{DMI_Ham} at different magnetic fields. The inset shows the used bipartition of the system in question.  (b) Mean and standard deviation of the entanglement entropy calculated for 22528 random eigenvectors superpositions generated with Haar measure in the range $h \in [0,0.28]$ as described in the text. (c) Distributions of the calculated entanglement entropy within the spin spiral phase.}
  \label{fig:flake_entropy}
\end{figure}%

We are now in a position to perform a preliminary analysis of  entanglement of the system being in the spin spiral phase with degenerate ground state for $0 \leq h \leq 0.28$. Fig.~\ref{fig:flake_entropy}\, (a) shows entanglement entropies for six eigenstates obtained from exact diagonalization of the spin model, Eq.\ref{DMI_Ham} at different magnetic fields. Fluctuations of the entropy values are significant and random. We do not observe any pattern in $\mathcal{S}_{\rm vN}$. Similar picture of data fluctuations when one calculating the topological entropy for a related model with degeneracy of the ground state was previously reported in Ref.\onlinecite{Levan}. The entropy variance becomes smaller for $h$ close to 0.2. 

Following the entanglement analysis of the Ising model presented above in the case of the DMI Hamiltonian we have generated 22528 samples of the degenerate eigenstates from six initial ones, $\Ket{\psi_{d}}$ ($d$ = 0,..,5) obtained from the exact diagonalization. In order to make thus generated eigenfunctions, $\Ket{\Psi} = \sum_{d} \alpha_{d} \Ket{\psi_d}$ uniformly distributed over the six-dimensional subspace spanned by $\Ket{\psi_d}$ vectors, the random complex coefficients are chosen according to the normal distribution. Figures~\ref{fig:flake_entropy} (b) and (c) demonstrate the entanglement behaviour of the degenerate ground states ensemble. The profile of the entanglement distribution (Fig.~\ref{fig:flake_entropy} (c)) is close to Gaussian one and differs from the Ising model result. The entropy averaged over the ensemble features the maximum at $h = 0.2$ and function discontinuity in the vicinity of $h = 0.28$. The latter can be explained by the changes in the degeneracy of the ground state. Overall, the behaviour of the von Neumann entropy $\mathcal{S}_{\rm vN} (\rho_{\rm A})$ for $h \in (0.2, 0.3)$ is complex and can be attributed to the transition of the quantum system to skyrmion state \cite{qskyrmion,Haller} at higher fields.

\begin{figure}
  \centering
  \includegraphics[width=0.93\linewidth]{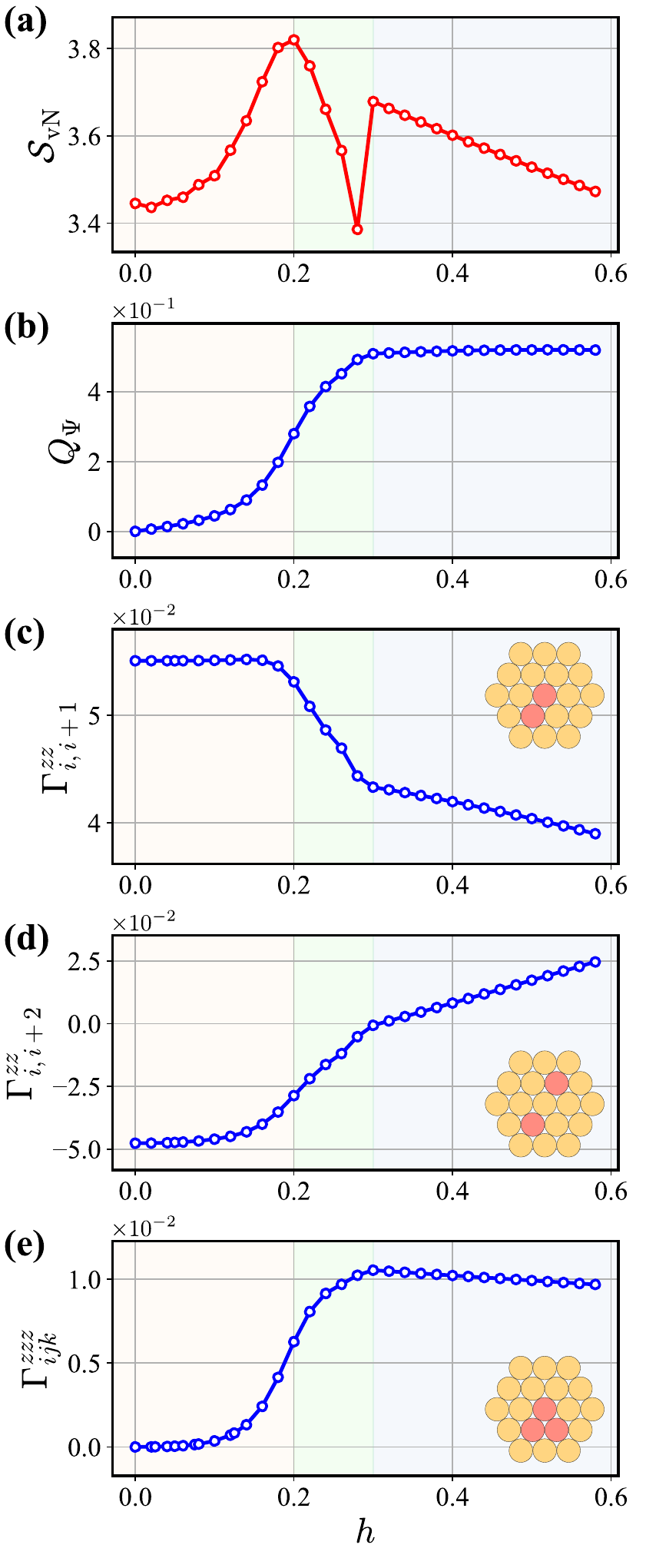}
  \caption{(a) Mean values of the entanglement entropy reproduced from Fig.\ref{fig:flake_entropy} (b). (b) Scalar chirality calculated with Eq.\ref{QPsi}. (c) and (d) Two-spin Ursell functions, Eq.\ref{two-spin} estimated for nearest and next-nearest neighbour spins, respectively. (e) Three-spin Ursell function calculated with Eq.\ref{three-spin}. Background colors denote spin-spiral (yellow), intermediate (green) and (blue) skyrmionic (blue) phases.}
  \label{fig:flake_entropy_and_chirality}
\end{figure}

To confirm the existence of the quantum skyrmion state in Fig.\ref{fig:flake_entropy_and_chirality} we reproduce the average zero-temperature value of the scalar chirality operator introduced in Ref.\cite{qskyrmion}
\begin{eqnarray}
Q_{\Psi} = \frac{N_{\triangle}}{\pi} \braket{\hat {\bf S}_{1} \, [\hat {\bf S}_{2} \times \hat {\bf S}_{3}]},
\label{QPsi}
\end{eqnarray}
where $N_{\triangle}$ is the number of non-overlapping triangles between nearest neighbours. The correlator in this equation is defined as  $\braket{\hat{X}} = \frac{1}{\mathcal{D}} \sum_{d} \braket{\psi_{d} |\hat{X}| \psi_{d}}$, where $\mathcal{D}$ is the degeneracy degree of the ground state and $\ket{\psi_{d}}$ is the eigenstate obtained with the exact diagonalization. 
The calculated $Q_{\Psi}$ is characterized by the plateau in the range $0.3\leq h < 0.68$, which is the signature of forming pure quantum skyrmion state. According to Ref.\cite{Entanglement_witnesses_chirality} the scalar chirality can be considered as a witness of tripartite entanglement in a quantum system, which calls for analyzing multi-partite correlations of the quantum skyrmion states. Below we consider the field dependence of another type of the three-spin correlation functions.

According to the results of previous investigations \cite{qskyrmion} at $h = 0$ the system in question is found to be in the pure spin-spiral state. There are six Bragg peaks at opposite wave vectors ${\bf q}$ and $-{\bf q}$ and zero chirality. In turn, in the range of the fields $0 < h < 0.28$  the system reveals properties inherent both quantum spin spiral and skyrmion phases. Interestingly, the peak of the entanglement entropy at $h =0.2$ (Fig.~\ref{fig:flake_entropy} (b)) evidences some changes in quantum correlations, which have not discussed in the previous works.

To elaborate this point we have analyzed Ursell spin-spin correlation functions~\cite{Ursell}
\begin{eqnarray}
\label{two-spin}
\Gamma_{ij}^{zz} = \braket{ S^z_{i} S^z_{j}}  - \braket{ S^z_{i} }\braket{S^z_{j}} 
\end{eqnarray}
between nearest and next-nearest neighbours as well as three-spin $zzz$ correlation function for an arbitrary triangle of neighbouring spins
\begin{eqnarray}
\label{three-spin}
\Gamma_{ijk}^{zzz} = \braket{S^z_{i} S^z_{j} S^{z}_k} - 3 \braket{S^z_{i}} \braket{S^z_{j} S^z_{k}}  + 2 \braket{S^z_{i}} \braket{S^z_{j}}\braket{S^z_{k}}. 
\end{eqnarray} 
The results of these calculations presented in Fig.\ref{fig:flake_entropy_and_chirality} show constant ferromagnetic behaviour of the pair correlation function for the nearest neighbour spins in the range of the fields $0 < h < 0.2$. For the fields higher than 0.2 we observe a suppression of $\Gamma_{ij}^{zz}$. As for the next-nearest neighbours correlations, they are  antiferromagnetic, demonstrate constant behaviour at $0\leq h < 0.15$ and are suppressed at higher magnetic fields. 
Interestingly, at the critical fields $h \in (0.28,0.3)$ corresponding to the transition from the spin spiral to the skyrmion phase, the $\Gamma_{ij}^{zz}$ correlator for next-nearest neighbours is zero.

Contrary to the pair Ursell functions, three-spin correlations ($\Gamma^{zzz}_{ijk}$) demonstrates the opposite trend, they amplify with increasing the magnetic field, which is similar to the behaviour of the scalar chirality discussed above. Such a similarity between $\Gamma^{zzz}_{ijk}$ and $Q_{\Psi}$ could be important from the experimental point of view. Here this is still an open question how to probe correlations beyond the one-and two point functions \cite{Entanglement_witnesses1} and, if possible, which correlation functions are easier to measure. In addition, the obtained magnetic field dependence of the correlation functions evidences that there is a smooth (continuous) transition between the spin spiral and skyrmion phases in the quantum case, which is completely different from the classical solution of the skyrmion problem \cite{qskyrmion}.

\section{Single-shot description of the degenerate systems} 
Having quantified entanglement of the ensembles of the degenerate eigenstates for some correlated systems with von Neumann entropy we are going to consider the characterization problem on the level of the projective measurements. 
Normally, in quantum computing or atomic simulations one prepares multiple copies of a given quantum state and performs their projective measurements \cite{Nielsen}. It allows to estimate correlation functions \cite{classical_shadow, RBM_Dicke}, entanglement entropy \cite{Heyl, Toric_code, classical_shadow} or other measures \cite{Sotnikov, Dicke_sign_structure, Japan_review, Hamming_nets, ourVQE} for describing the system in question. In the degenerate case we consider multiple copies of the same eigenstate may not be available in real experiments \cite{Rydberg1, Ising_dissipation}. That is why it is instructive to imitate an extreme situation when every eigenstate that is prepared for measurement belongs to the degenerate manifold and is measured only once. A finite set of the bitstrings obtained within such a single-shot protocol contains some information on the degenerate manifold of the ground state. However, the usefulness of these data is not a priori obvious and requires additional study. 

\begin{figure}
  \centering
  \includegraphics[width=1\linewidth]{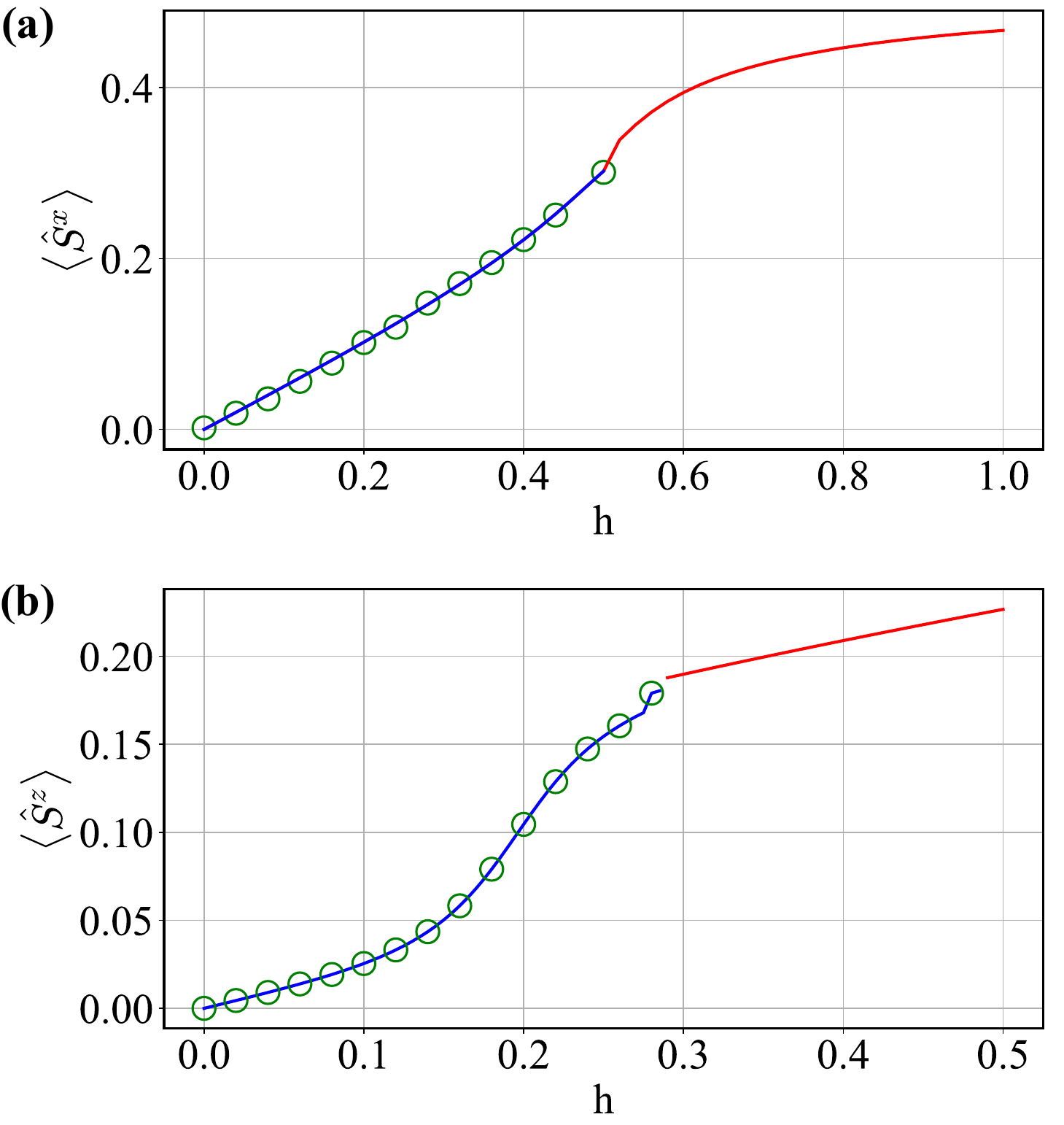}
  \caption{Calculated magnetizations and the corresponding  approximations obtained within single-shot protocol for (a) Ising model and (b) DMI Hamiltonian. Red lines denote the magnetization for non-degenerate ground state taken from exact diagonalization. Blue lines stand for results obtained with 2-fold (Ising model) and 6-fold (DMI model) degenerate eigenfunctions. Circles correspond to the magnetization values, Eq.\ref{Haar_mag} averaged over ensemble of the degenerate eigenstates by using single-shot protocol described in the text. }
  \label{magnetization}
\end{figure}

In our work we focus on estimating the local correlation function in the ground state that can be formally written as
\begin{eqnarray}
\langle {\hat S}^{\mu}_{i} \rangle = \frac{1}{\mathcal{D}} \sum_{d} \braket{\psi_{d} | {\hat S}^{\mu}_{i} | \psi_{d} },
\end{eqnarray}
where the index $d$ denotes degenerate eigenstates, $\mu$ stands for projection of the spin operator and $i$ is the site index. For the Ising and DMI models we consider the local magnetization can be calculated exactly. In Fig.\ref{magnetization} we present these exact values of the local magnetization with lines. Blue color denotes the range of the magnetic fields with degenerate solutions. These results fully agree with that known from the literature \cite{Pfeuty, qskyrmion}.

We are now in a position to define an analog of the local correlation function that can be estimated from the single-shot protocol for degenerate eigenstates described above. More specifically, for each site we are to estimate the following quantity
\begin{eqnarray}
\langle {\hat S}^{\mu} \rangle_{\rm Haar} = \frac{p(\uparrow) - p(\downarrow)}{2},
\label{Haar_mag}
\end{eqnarray}
where $p(\uparrow)$ ($p(\downarrow)$) is the probability to get the spin-up (spin-down) state for the given site when performing measurements. This quantity is estimated for a finite set of the degenerate eigenstates superpositions generated with the normal distribution law. Practically, the probability function is estimated from the finite set of bitstrings obtained with single-shot protocol. The number of the bitstrings is equal to 8192 and 256 for the Ising and DMI models, respectively. The basis for measurements is defined by the direction of the external magnetic field in the considered model ($x$ for Ising Hamiltonian and $z$ for the DMI model). The calculated values of $\langle {\hat S}^{\mu} \rangle_{\rm Haar}$ averaged over spins are presented in Fig.\ref{magnetization} with circles. One can see that the single-shot protocol provides reliable approximation of the exact data and can be used in real experiments dealing with degenerate quantum systems.

\section{Simple fermionic and bosonic systems with degeneracy of the ground state}
To complete our study of the quantum systems with degenerate ground state in this section we extend our consideration onto systems interacting fermions or bosons that attract considerable attention in modern experiments on ultra-cold atoms \cite{Ultracold}. As we will demonstrate the simplest two-site Fermi-Hubbard and Bose-Hubbard models can reveal interesting entanglement properties in some limiting cases.

First we discuss the results obtained for two-site fermionic Hubbard model \cite{Hubbard} that is given by
\begin{eqnarray}
H_{\rm FH} = - t \sum_{i j \sigma} c^{\dagger}_{i\sigma} c_{j \sigma} + U \sum_{i} n_{i \uparrow} n_{i \downarrow} 
- \mu \sum_{i \sigma} n_{i \sigma}, 
\end{eqnarray}
where $t_{ij}$ is the hopping integral, $U$ stands for the on-site Coulomb interaction, $\mu$ is chemical potential. $c^{\dagger}_{i \sigma}$ ($c_{i \sigma}$) and $n_{i \sigma}$ are the fermionic creation (annihilation) and particle number operators, respectively. Each site can be in four different states: empty $\ket{0}$, occupied by an electron with spin up $\ket{\uparrow}$, occupied by an electron with spin down $\ket{\downarrow}$ and occupied by two electrons with opposite spins $\ket{\uparrow \downarrow}$.

Among possible solutions of the Hubbard model we are primarily interested in exploring those with the degeneracy of the ground state. For instance, this is the case when there is single electron in two-site system, whose ground state is described by the following eigenstates
\begin{eqnarray}
\ket{\psi_{\uparrow}} = \frac{1}{\sqrt{2}} (\ket{\uparrow 0} + \ket{0 \uparrow}), \\
\ket{\psi_{\downarrow}} = \frac{1}{\sqrt{2}} (\ket{\downarrow 0} + \ket{0 \downarrow}). 
\end{eqnarray}
The corresponding ground state energy is equal to $-t -\mu$.
For single-site subsystem the entanglement entropy of each state is equal to 1 and relates to our uncertainty to find the electron with spin up or spin down at the specific site. In turn, the von Neumann entropy calculated for an arbitrary superposition of these degenerate wave functions, $\alpha_0 \ket{\psi_{\uparrow}} + \alpha_1 \ket{\psi_{\downarrow}}$ is likewise equal to 1 independent on the choice of the coefficients. Thus, the degeneracy relating to the spin channels doesn't affect quantum correlation between different sites in the considered single-particle case.

Next we are going to discuss the two-site system of interacting bosons that are described with the following Hubbard model
\begin{eqnarray}
\label{Bose_Ham}
H_{\rm BH} = - t \sum_{i j} a^{\dagger}_{i} a_{j} + \frac{U}{2} \sum_{i} n_{i} (n_{i} - 1)
- \mu \sum_{i} n_{i},
\end{eqnarray}
where $a^{\dagger}_{i}$ ($a_{i}$) and $n_{i}$ are the bosonic creation (annihilation) and particle number operators, respectively.

To simplify our search for the degenerate ground states in the two-site bosonic systems it is instructive to consider the insulating regime when the hopping integral in Eq.\ref{Bose_Ham} is zero. In this case the Hamiltonian matrix of the $N$-boson system is diagonal in the Fock basis $\{ \ket{n} \} = \{ \ket{0,N}, \ket{1,N-1} ... \ket{N,0} \}$, here $n$ encodes the occupation of the first site \cite{Grabovsky_thesis}. The corresponding eigenvalues are expressed as $E_{n} = U(n^2 - Nn) + \frac{U}{2} N^2 - (\frac{U}{2} + \mu)N$.

For $N$=1 there is two-fold degenerate ground state with $E_{0} =E_{1} = -\mu$ and the corresponding wave functions are
$\ket{\psi_0} = \frac{1}{\sqrt{2}} (\ket{01} + \ket{10})$ and $\ket{\psi_1} = \frac{1}{\sqrt{2}} (\ket{01} - \ket{10})$. While each of these wave functions features 50\% probability to find the boson at the specific site and the maximal entanglement entropy value of 1, this is not the case for their superpositions within the degeneracy manifold. For instance, the combination $\frac{1}{\sqrt{2}}(\ket{\psi_0} + \ket{\psi_{1}})$ gives zero entanglement, since the boson occupies the first site. The distribution of the entanglement is almost the same as found in the case of the Ising model Fig.\ref{fig:ising_entanglement} a.

If the total number of bosons is larger than 1 and is even, the system's ground state is represented by a simple tensor product of the local Fock states with exactly $\frac{N}{2}$ bosons per site \cite{Ultracold}. There is no degeneracy of the ground state in this case. In turn, if $N$ is odd, the ground state is two-fold degenerate and the entanglement properties are similar to the $N$=1 case.

\section{Conclusion}
In this work, we address the problem of characterizing quantum systems in degenerate ground states with a special focus on the properties of entanglement entropy. On the basis of the linearly independent degenerate eigenstates calculated with exact diagonalization we propose to generate a finite ensemble of their linear combinations with Haar measure. This allows to probe the entanglement distributions of the ground wave functions. The primers we consider in this work show that such distributions are problem specific. For instance, in the case of the transverse-field Ising model the calculated values of the von Neumann entropy are non-uniformly distributed in the wide range between 0 and 1. At the same time, for the quantum system with non-collinear order the entanglement distribution reveals a more localized Gaussian-like shape. The found enhancement of the mean entropy of the degenerate eigenstate ensemble in the spin spiral phase can be explained by the changes in three- and two-spin correlations which demonstrate different trends at increasing the external magnetic field. To establish a connection between our theoretical results and real experiments we analyze the important details concerning estimation of the observables in the case of the degenerate quantum systems.   

\section*{Acknowledgments}
We thank Stanislav Straupe, Aleksey Fedorov and Evgeniy Kiktenko for fruitful discussions. The research funding from the Ministry of Science and Higher Education of the Russian Federation (Ural Federal University Program of Development within the Priority-2030 Program, project 4.72) is gratefully acknowledged. The quantum VQE simulations were performed within the Russian Roadmap on Quantum Computing (Contract No. 868-1.3-15/15-2021, October 5, 2021) in the development of the trapped-ion processor. The investigations of the fermionic and bosonic models were carried out with the financial support of the Ministry of Science and Higher Education of the Russian Federation (theme FEUZ-2023-0013). Quantum simulations were performed on the Uran supercomputer at the IMM UB RAS.


\begin{thebibliography}{0}

\bibitem{Ising_dissipation}
X. Mi et al., Stable quantum-correlated many-body states through engineered dissipation, Science 383, 1332–1337 (2024).

\bibitem{Heyl}
Jun Yong Khoo and Markus Heyl, Quantum entanglement recognition, Phys. Rev. Research 3, 033135 (2021)

\bibitem{Carleo}
G. Carleo and M. Troyer, M. Solving the quantum many-body problem with artificial neural networks. Science 355, 602–606 (2017).

\bibitem{Rydberg1}
Pascal Scholl, Michael Schuler, Hannah J. Williams, Alexander A. Eberharter, Daniel Barredo, Kai-Niklas Schymik, Vincent Lienhard, Louis-Paul Henry, Thomas C. Lang, Thierry Lahaye, Andreas M. L\"auchli and Antoine Browaeys, Quantum simulation of 2D antiferromagnets with hundreds of Rydberg atoms, Nature 595, 233 (2021) 

\bibitem{Rydberg2}
P. Schauss, J. Zeiher, T. Fukuhara, S. Hild, M. Cheneau, T. Macr\`{i}, T. Pohl, I. Bloch, C. Gross, Crystallization in Ising
quantum magnets, Science 347, 1455 (2015).

\bibitem{IBM127}
Youngseok Kim, Andrew Eddins, Sajant Anand, Ken Xuan Wei, Ewout van den Berg, Sami Rosenblatt, Hasan Nayfeh, Yantao Wu, Michael Zaletel, Kristan Temme and Abhinav Kandala, Evidence for the utility of quantum computing before fault tolerance,  Nature 618, 500 (2023).

\bibitem{Toric_code}
K. J. Satzinger, Y. Liu, A. Smith, C. Knapp, M. Newman, C. Jones, Z. Chen, C. Quintana, X. Mi, A. Dunsworth, C. Gidney, I. Aleiner, F. Arute, K. Arya, J. Atalaya, R. Babbush, J. C. Bardin, R. Barends, J. Basso, A. Bengtsson, A. Bilmes, M. Broughton, B. B. Buckley, D. A. Buell, B. Burkett, N. Bushnell, B. Chiaro, R. Collins, W. Courtney, S. Demura, A. R. Derk, D. Eppens, C. Erickson, E. Farhi, L. Foaro, A. G. Fowler, B. Foxen, M. Giustina, A. Greene, J. A. Gross, M. P. Harrigan, S. D. Harrington, J. Hilton, S. Hong, T. Huang, W. J. Huggins, L. B. Ioffe, S. V. Isakov, E. Jeffrey, Z. Jiang, D. Kafri, K. Kechedzhi, T. Khattar, S. Kim, P. V. Klimov, A.N. Korotkov, F. Kostritsa, D. Landhuis, P. Laptev, A. Locharla, E. Lucero, O. Martin, J. R. McClean, M. McEwen, K. C. Miao, M. Mohseni, S. Montazeri, W. Mruczkiewicz, J. Mutus, O. Naaman, M. Neeley, C. Neill, M. Y. Niu, T. E. O'Brien, A. Opremcak, B. Pató, A. Petukhov, N. C. Rubin, D. Sank, V. Shvarts, D. Strain, M. Szalay, B. Villalonga, T. C. White, Z. Yao, P. Yeh, J. Yoo, A. Zalcman, H. Neven, S. Boixo, A. Megrant, Y. Chen, J. Kelly, V. Smelyanskiy, A. Kitaev, M. Knap, F. Pollmann, P. Roushan,
Realizing topologically ordered states on a quantum processor, Science 374, 1237-1241 (2021).

\bibitem{Anderson_towers}
P. W. Anderson, An approximate quantum theory of the antiferromagnetic ground state, Phys. Rev. 86, 694 (1952).

\bibitem{Lauchli_towers}
A. Wietek, M. Schuler, and A.M. L\"auchli, Studying continuous symmetry breaking using energy level spectroscopy, arXiv:1704.08622.

\bibitem{Quantum_Darwinism}
O.M. Sotnikov, E.A. Stepanov, M.I. Katsnelson, F. Mila, V.V. Mazurenko, Emergence of Classical Magnetic Order from Anderson Towers: Quantum Darwinism in Action, Physical Review X 13, 041027 (2023).

\bibitem{Lauchli1}
A. M. L\"auchli, J. Sudan, and E. S. S\o rensen, Ground-state energy and spin gap of spin-1 kagome-Heisenberg antiferromagnetic clusters: Large-scale exact diagonalization results, Phys. Rev. B 83, 212401 (2011).

\bibitem{Lauchli2}
A. Wietek and A. M. L\"auchli, Sublattice coding algorithm and distributed memory parallelization for large-scale exact diagonalizations of quantum many-body systems, Phys. Rev. E 98, 033309 (2018).

\bibitem{Fehske}
Alexander Weisse and Holger Fehske, Exact Diagonalization Techniques, Lecture Notes in Physics 739, 529 (2008).

\bibitem{Manmana}
R.M. Noack and S. Manmana, Diagonalization- and Numerical Renormalization-Group-Based Methods for Interacting Quantum Systems, AIP Conf. Proc. 789, 93 (2005)

\bibitem{Joshi}
Ashish Joshi, Robert Peters, and Thore Posske, Ground state properties of quantum skyrmions described by neural network quantum states, Phys. Rev. B 108, 094410 (2023).

\bibitem{Bagrov1}
T. Westerhout, N. Astrakhantsev, K. S. Tikhonov, M. I. Katsnelson, and A. A. Bagrov, Generalization properties of neural network approximations to frustrated magnet ground states, Nat. Commun. 11, 1593 (2020)

\bibitem{Bagrov2}
Tom Westerhout, Mikhail I. Katsnelson, Andrey A. Bagrov, Many-body quantum sign structures as non-glassy Ising models, Communications Physics 6, 275 (2023).

\bibitem{Sotnikov}
O.M. Sotnikov, I.A. Iakovlev, A. A. Iliasov, M. I. Katsnelson, A. A. Bagrov, V.V. Mazurenko, Certification of quantum states with hidden structure of their bitstrings, npj Quantum Information 8, 41 (2022).

\bibitem{Popkov1}
Vladislav Popkov, Mario Salerno, Logarithmic divergence of the block entanglement entropy for the ferromagnetic Heisenberg model, Phys. Rev. A 71, 012301(2005).

\bibitem{Popkov2}
Vladislav Popkov, Mario Salerno, Gunter Schuetz, Entangling power of permutation invariant quantum states, 	Phys. Rev. E72, 032327 (2005).

\bibitem{Doyon1}
Olalla A. Castro-Alvaredo, Benjamin Doyon, Permutation operators, entanglement entropy, and the XXZ spin chain in the limit $\Delta \rightarrow -1$, 	J.Stat.Mech. 1102, P02001 (2011).

\bibitem{Doyon2}
Olalla A. Castro-Alvaredo, Benjamin Doyon, Entanglement entropy of highly degenerate states and fractal dimensions, Phys. Rev. Lett. 108, 120401 (2012).

\bibitem{Doyon3}
Olalla A. Castro-Alvaredo, Benjamin Doyon, Entanglement in permutation symmetric states, fractal dimensions, and geometric quantum mechanics, 	J. Stat. Mech. P02016 (2013).

\bibitem{Pfeuty}
P. Pfeuty, The one-dimensional Ising model with a transverse field, Ann. Phys. (N.Y.) 57, 79 (1970)

\bibitem{Fernandez}
P. Q. Cruz, G. Catarina, R. Gautier, J. Fern\'andez-Rossier, Optimizing quantum phase estimation for the simulation of Hamiltonian eigenstates, Quantum Sci. Technol. 5 044005 (2020).

\bibitem{Ising_QAOA}
Wen Wei Ho, Timothy H. Hsieh, Efficient variational simulation of non-trivial quantum states, SciPost Phys. 6, 029 (2019).

\bibitem{DTC_PRX}
Matteo Ippoliti, Kostyantyn Kechedzhi, Roderich Moessner, S.L. Sondhi, and Vedika Khemani, Many-Body Physics in the NISQ Era: Quantum Programming a Discrete Time Crystal, PRX QUANTUM 2, 030346 (2021)

\bibitem{DTC_Google}
X. Mi et al., Time-crystalline eigenstate order on a quantum processor, Nature (London) 601, 531 (2022).

\bibitem{ourDTC}
E. A. Maletskii, I. A. Iakovlev, and V. V. Mazurenko, Quantifying spatiotemporal patterns in classical and quantum systems out of equilibrium, Phys. Rev. E 109, 024105 (2024)

\bibitem{Tom}
T. Westerhout, Lattice-symmetries: A package for working with quantum many-body bases, J. Open Source Softwaare 6, 3537 (2021).

\bibitem{Rydberg}
H. Bernien, S. Schwartz, A. Keesling, H. Levine, A. Omran, H. Pichler, S. Choi, A.S. Zibrov, M. Endres, M. Greiner, V. Vuleti\'c and M.D. Lukin, Probing many-body dynamics on a 51-atom quantum simulator, Nature 551, 579 (2017).

\bibitem{ions}
C. Monroe, W.C. Campbell, L.-M. Duan, Z.-X. Gong, A.V. Gorshkov, P.W. Hess, R. Islam, K. Kim, N.M. Linke, G. Pagano, P. Richerme, C. Senko, and N.Y. Yao, Programmable quantum simulations of spin systems with trapped ions, Rev. Mod. Phys. 93, 025001 (2021).

\bibitem{entropy_Ising1}
J. I. Latorre, E. Rico and G. Vidal, Ground state entanglement in quantum spin chains, Quantum Inf. Comput. 4, 048 (2004)

\bibitem{entropy_Ising2}
P. Calabrese and J. Cardy, Entanglement entropy and quantum  field theory, Journal of Statistical Mechanics: Theory and Experiment 2004(06), P06002 (2004)

\bibitem{DMRG}
I. Peschel, M. Kaulke, Ö. Legeza, Density-matrix spectra for integrable models, Ann. Physik (Leipzig) 8, 153 (1999).

\bibitem{Haar_qiskit}
Karol Zyczkowski and Hans-J\"urgen Sommers, Induced measures in the space of mixed quantum states, Journal of Physics A: Mathematical and General 34, 7111 (2001).

\bibitem{Entanglement_witnesses1}
P. Laurell, A. Scheie, E. Dagotto, D. A. Tennant, Witnessing Entanglement and Quantum Correlations in Condensed Matter: A Review, Adv. Quantum Technol. 2400196 (2024).

\bibitem{Entanglement_witnesses2}
Allen Scheie, Pontus Laurell, Wolfgang Simeth, Elbio Dagotto, and D. Alan Tennant, Tutorial: Extracting entanglement signatures from neutron spectroscopy, Materials Today Quantum 5, 100020 (2025).

\bibitem{VQE}
Abhinav Kandala, Antonio Mezzacapo, Kristan Temme, Maika Takita, Markus Brink, Jerry M. Chow and Jay M. Gambetta, Hardware-efficient variational quantum eigensolver for small molecules and quantum magnets, Nature \textbf{549}, 242 (2017).

\bibitem{qiskit}
qiskit.org

\bibitem{Joos1}
E. Joos, H. D. Zeh, C. Kiefer, D. J.W. Giulini, J. Kupsch, and I. O. Stamatescu, \textit{Decoherence and the Appearance of a Classical World in Quantum Theory} (Springer, Berlin, 2013).

\bibitem{Zurek0} 
W. H. Zurek, ``Pointer basis of quantum apparatus: Into what mixture does the wave packet collapse?'' Phys. Rev. D \textbf{24}, 1516 (1981).

\bibitem{Zurek1}
W. H. Zurek, ``Decoherence and the Transition from Quantum to Classical,'' Physics Today \textbf{44}, 10, 36 (1991).

\bibitem{Zurek3}
W. H. Zurek, ``Decoherence, einselection, and the quantum origins of the classical,'' Rev. Mod. Phys. \textbf{75}, 715 (2003).

\bibitem{Zurek2}
W. H. Zurek, ``Quantum Darwinism,'' Nat. Phys. \textbf{5}, 181 (2009).

\bibitem{Dicke}
R. H. Dicke, Coherence in Spontaneous Radiation Processes, Physical Review 93(1), 99 (1954).

\bibitem{Dicke_correlations}
Davide Girolami, Tommaso Tufarelli, and Cristian E. Susa, Quantifying Genuine Multipartite Correlations and their Pattern Complexity, PRL 119, 140505 (2017).


\bibitem{qskyrmion}
O.M. Sotnikov, V.V. Mazurenko, J. Colbois, F. Mila, M.I. Katsnelson, and E.A. Stepanov, Probing the topology of the quantum analog of a classical skyrmion,
Phys. Rev. B \textbf{103}, L060404 (2021)

\bibitem{Levan}
Vipin Vijayan, L. Chotorlishvili, A. Ernst, S. S. P. Parkin, M. I. Katsnelson, S. K. Mishra, Topological entanglement entropy to identify topological order in quantum skyrmions, arXiv:2311.03915.

\bibitem{Haller}
Andreas Haller, Solofo Groenendijk, Alireza Habibi, Andreas Michels, and Thomas L. Schmidt, Quantum skyrmion lattices in Heisenberg ferromagnets, Phys. Rev. Research 4, 043113 (2022).

\bibitem{Entanglement_witnesses_chirality}
Dimitris I. Tsomokos, Juan Jos\'e Garc\'ia Ripoll, Nigel R. Cooper, and Jiannis K. Pachos, Chiral entanglement in triangular lattice models, Physical Review A 77, 012106 (2008).  

\bibitem{Ursell}
H.D. Ursell, The evaluation of Gibbs' phase-integral for imperfect gases, in Mathematical Proceedings of the Cambridge Philosophical Society, Vol. 23 (Cambridge University Press, 1927) pp. 685–697.

\bibitem{Nielsen}
Michael A. Nielsen and Isaac L. Chuang, Quantum Computation and Quantum Information (Cambridge University Press, 2000)

\bibitem{classical_shadow}
H.-Y. Huang, R. Kueng, and J. Preskill, Predicting many properties of a quantum system from very few measurements. Nat. Phys. 16, 1050–1057 (2020).

\bibitem{RBM_Dicke}
Oleg M. Sotnikov, Ilia A. Iakovlev, Evgeniy O. Kiktenko, Mikhail I. Katsnelson, Aleksey K. Fedorov, Vladimir V. Mazurenko, Emergence of global receptive fields capturing multipartite quantum correlations, arXiv: 2408.13033.

\bibitem{Dicke_sign_structure}
Oleg M. Sotnikov, Ilia A. Iakovlev, Evgeniy O. Kiktenko, Aleksey K. Fedorov, Vladimir V. Mazurenko, Achieving the volume-law entropy regime with random-sign Dicke states, Physical Review A 110, 062416 (2024).

\bibitem{Japan_review}
Vladimir V Mazurenko, Ilia A Iakovlev, Oleg M Sotnikov, Mikhail I Katsnelson, Estimating patterns of classical and quantum skyrmion states, Journal of the Physical Society of Japan 92, 081004 (2023).

\bibitem{Hamming_nets}
Ilia A. Iakovlev, Oleg M. Sotnikov, Ivan V. Dyakonov, Evgeniy O. Kiktenko, Aleksey K. Fedorov, Stanislav S. Straupe, Vladimir V. Mazurenko, Benchmarking a boson sampler with Hamming nets, Phys. Rev. A 108, 062420 (2023).

\bibitem{ourVQE}
O.M. Sotnikov and V.V. Mazurenko, Neural network agent playing spin Hamiltonian games on a quantum computer, J. Phys. A: Math. Theor. 53 135303 (2020).

\bibitem{Ultracold}
Immanuel Bloch, Jean Dalibard, Wilhelm Zwerger, Many-Body Physics with Ultracold Gases, Rev. Mod. Phys. 80, 885 (2008).

\bibitem{Hubbard}
J. Hubbard, Electron correlations in narrow energy bands, Proc. R. Soc. London, Ser. A 276, 238 (1963).

\bibitem{Grabovsky_thesis}
David Grabovsky, The limits of the Hubbard model, undergraduate thesis (2019).


\end{thebibliography}
\end{document}